# A new 2D auxetic $CN_2$ nanostructure with high energy density and mechanical strength

Qun Wei[1], Ying Yang[2,3], Alexander Gavrilov[2], Xihong Peng[2*]

[1]School of Physics and Optoelectronic Engineering, Xidian University, Xi'an, Shaanxi 710071, P. R. China

[2]College of Integrative Sciences and Arts, Arizona State University, Mesa, Arizona 85212, USA

[3]School of Automation and Information Engineering, Xi'an University of Technology, Xi'an, Shaanxi, 710048, P. R. China

## ABSTRACT

The existence of a new two dimensional $CN_2$ structure was predicted using *ab-initio* molecular dynamics (AIMD) and density-functional theory calculations. It consists tetragonal and hexagonal rings with C-N and N-N bonds arranged in a buckling plane, isostructural to tetrahex-carbon allotrope. It is thermodynamically and kinetically stable suggested by its phonon spectrum and AIMD. This nanosheet has high concentration of N and contains N-N single bonds with an energy density of 6.3 kJ/g, indicating potential applications as high energy density materials. It possesses exotic mechanical properties with negative Poisson's ratio and an anisotropic Young's modulus. The modulus in the zigzag direction is predicted to be 340 N/m, stiffer than *h*-BN and penta-$CN_2$ sheets and comparable to graphene. Its ideal strength of 28.8 N/m outperforms that of penta-graphene. The material maintains phonon stability upon the application of uniaxial strain up to 10% (13%) in the zigzag (armchair) direction or biaxial strain up to 5%. It possesses a wide indirect HSE band gap of 4.57 eV which is tunable between 3.37 eV ~ 4.57 eV through strain. Double-layer structures are also explored. Such unique properties may have potential applications in high energy density materials, nanomechanics and electronics.

---

[*] To whom correspondence should be addressed. E-mail: xihong.peng@asu.edu, Phone: 1-480-727-5013.

## 1. Introduction

Fruitful development and synthesis of two dimensional (2D) materials such as graphene [1–3], transition metal dichalcogenides (TMDs) [4–7], MXenes [8], and phosphorene [9,10], prompt incredible research interests in 2D materials. These already fabricated 2D materials have shown promising applications in photovoltaic [11], photocatalytic [12], transistors [13], photodetectors [14], superconductivity [15], and batteries [16], which stimulates further discovery and investigation of new 2D materials with unique properties. Among them, 2D carbon nitride structures draw substantial attention due to their distinctive strong covalent bond network and rich physical and chemical properties.

A variety of 2D carbon nitride structures have already been successfully fabricated in labs in the past years. For example, graphitic carbon nitride (g-$C_3N_4$) [17] can be made via polymerization-type synthesis and demonstrates unusual catalytic activity as a metal-free catalysis for many chemical reactions [17]. The nitrogenated holey graphene called $C_2N$-$h$2D crystal structure [18] can be efficiently synthesized via wet-chemical reaction and has great semiconducting properties with the band gap in the visible range. The 2D polyaniline structure with $C_3N$ stoichiometry [19] has been realized by direct solid-state reaction of organic single crystals and exhibits excellent conductivity. Graphite-like lamellar structure g-CN [20] was also reported to be synthesized via a solvent-free route at low temperatures. In addition to experimental investigation, numerous 2D carbon nitrides were also theoretical predicted, including $C_{12}N$ [21], g-$C_4N_3$ [22], $C_{10}N_9$ [23], $C_6N_6$ [23–25], and $C_6N_8$ [23,25] *etc*.

Nitride materials may have potential application as high energy density materials (HEDM). High concentration of nitrogen is much desired for such application. The above mentioned 2D carbon nitrides have relatively low nitrogen to carbon ratio with 4/3 being the highest in g-$C_3N_4$. Zhang *et al.*, theoretically predicted a carbon nitride nanosheet named as penta-$CN_2$ [26], which has much higher nitrogen content and was reported to have an energy density of 4.41 kJ/g [26].

In this work, we predicted a new 2D carbon nitride nanosheet with also $CN_2$ stoichiometry and high energy density of 6.3 kJ/g, higher than that in penta-$CN_2$ and the nitrogen rich B-N compound [27]. In the structure, the $sp^3$ hybridized carbon atoms were sandwiched between two layers of nitrogen atoms in a buckling plane, isostructural to tetrahex-C [28]. Since this structure contains tetragonal (T) and hexagonal (H) rings, it is named as TH-$CN_2$. The structure is proven to be thermodynamically stable, suggested by the phonon spectrum and *ab initio* molecular dynamics (AIMD) calculations.

This newly predicted TH-CN$_2$ structure shows lots of other merits. It has an in-plane axial Young's modulus of 340 N/m, stiffer than the *h*-BN monolayer [29] and penta-CN$_2$ [26], and comparable to graphene (348 N/m). It possesses a band gap of 4.57 eV suggesting insulating nature. In addition, this new material exhibits auxetic property with negative Poisson's ratio. It is also found that this thermodynamically stable structure proven to maintain stable with the application of mechanical strain. The phonon instability occurs with the uniaxial strain beyond 10% (13%) in the zigzag (armchair) direction or biaxial strain larger than 5%.

In this paper, we report our findings on the structure, stability, mechanical and electronic properties, double-layer configurations of the new 2D CN$_2$ nanosheet through the calculations of first-principles density-functional theory (DFT) [30] and AIMD.

## 2. Computational details

The first-principles DFT [30] calculations are performed using the VASP package [31,32] with the projector-augmented wave (PAW) potentials [33,34]. The Perdew-Burke-Ernzerhof (PBE) exchange-correlation functional [35] is chosen for general geometry relaxation and mechanical property calculations. The hybrid Heyd-Scuseria-Ernzerhof (HSE)06 method [36,37] is used to calculate electronic band structure and band gap because of its better performance on predicting semiconductor band gaps. For double-layer structures, in addition to PBE, we use two additional functionals PBEsol [38] and DFT-D3 [39] which provide an improved description of crystals including Van der Waals interaction.

The wave functions of valence electrons are described using the plane wave basis set with kinetic energy cutoff 900 eV. The reciprocal space is meshed 15 × 13 × 1 using Monkhorst-Pack method [40]. The energy convergence criterion for electronic iterations is set to be $10^{-6}$ eV and the force is converged to be less than 0.001 eV/ Å for geometry optimization of the simulation cell. The kinetic energy cutoff 500 eV for plane wave basis set is used for the HSE band structure calculations. In the band structure, 11 *k*-points are collected along each high symmetry line in the reciprocal space. The *c*-vector of the unit cell is set to be 20 Å to ensure sufficient vacuum space (> 16 Å) included in the calculations to minimize the interaction between the system and its replicas resulted from periodic boundary conditions. Phonon spectrum is calculated using a supercell approach in the PHONOPY code [41] with the forces computed from VASP [31,32]. Structural snapshots and electron orbital contour plots are generated using VESTA program [42].

### 3. Results and discussion

### 3.1. Structure and stability of TH-CN$_2$

The crystal structure of the 2D TH-CN$_2$ is given in Fig. 1. It is a buckled three-sublayer crystal structure with four-coordinated carbon atoms being sandwiched between two sublayers of three-coordinated nitrogen atoms. The rectangular conventional cell contains four C and eight N atoms. The lattice constants for the conventional cell from our calculations are $a = 4.18$ Å and $b = 5.78$ Å. The buckling thickness $d$ between the two N sublayers is 1.48 Å. The C-N bond $r_1$ and the N-N bond $r_2$ are predicted to be 1.47 Å and 1.46 Å, respectively. The calculated bond angles denoted in Fig. 1(a) are $\alpha = 121.8°$, $\beta = 119.1°$, and $\gamma = 90.9°$.

It is clear from the tilt-axis view in Fig. 1(d), two neighbored hexagonal rings along the **a**-axis are not coplanar. The dihedral angle between these two hexagonal rings is denoted as $\phi_{1234}$ by the neighboring atoms 1-2-3-4 and it is predicted to be $\phi_{1234} = 109.3°$. Similarly, the neighbored hexagonal and tetragonal rings are neither coplanar and its dihedral angle $\phi_{2345} = 124.4°$.

A summary of the structural parameters of TH-CN$_2$, along with penta-CN$_2$ [26], tetrahex-carbon [28,43], penta-graphene [44], and graphene is given in Table 1. Comparing TH-CN$_2$ with its isostructural carbon allotrope (tetrahex-C), it is found that the lattice constants $a$ and $b$ are reduced by 7.8% and 5.3%, respectively, while the buckling thickness $d$ is increased by 27.4%, indicating the structure of TH-CN$_2$ is more buckled. The bond length $r_1$ describing the bonds between the $sp^3$ and $sp^2$ carbon atoms in tetrahex-C is now representing the C-N bond in TH-CN$_2$ and its value is reduced by 4.5%. However, the length $r_2$ for the bond between two $sp^2$ carbon atoms in tetrahex-C is now for the N-N bond and it lengthens by 9.4%. The lengthening of $r_2$ is mainly resulted from their different bond nature. The C-C bond $r_2$ in tetrahex-C is doubly bonded while N-N in TH-CN$_2$ is a single bond. The changes in bond angles are also observed, +8.5%, -3.9% and -4.5% for the angles $\alpha, \beta$ and $\gamma$, respectively. The dihedral angles are both decreased by 13% and 9.4% for $\phi_{1234}$ and $\phi_{2345}$, respectively. Similar structural variation trends are also observed when comparing penta-CN$_2$ with penta-graphene.

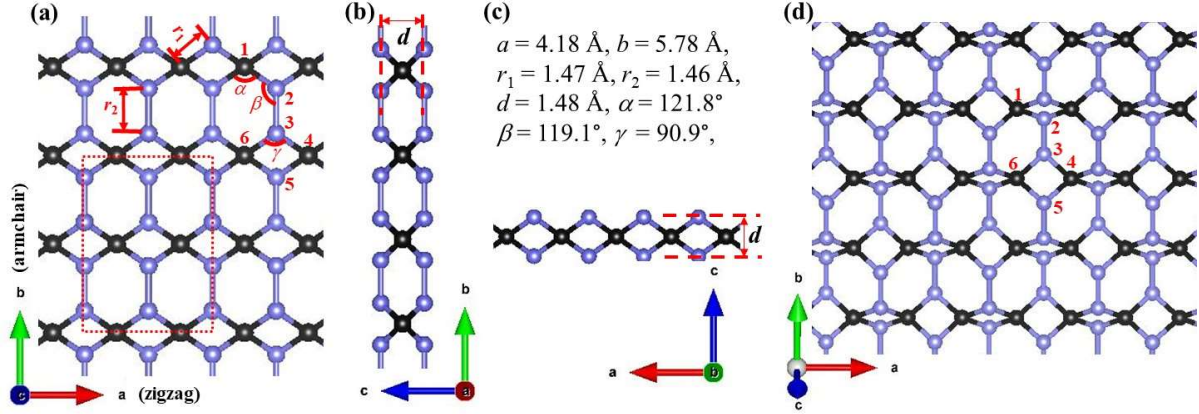

Figure 1. Snapshots of the 2D TH-CN$_2$ structure. The dashed rectangle in (a) represents a conventional cell. The carbon and nitrogen atoms are in black and blue, respectively. The buckling thickness d of the structure is labeled in (b) and (c). The lattice constants, bond lengths, buckling thickness, bond angles are given in (c). (d) Tilted-axis view of the TH-CN$_2$ structure.

Table 1. Summary of structural parameters and basic properties of TH-CN$_2$, tetrahex-C, penta-CN$_2$, penta-graphene, and graphene. Lattice constants a, b, buckling thickness d, and bond lengths $r_1$, $r_2$ are in unit of Å, cohesive and formation energies are in eV/atom, energy density $\rho$ in kJ/g, modulus $E_x$ in N/m, HSE predicted band gap $E_g$ in eV. The formation energy is calculated using graphene and molecular nitrogen as references (thus *zero for graphene). Reference: tetrahex-C [28,43], penta-CN$_2$ [26], penta-graphene [44].

| Structures | a | b | d | $r_1$ | $r_2$ | $E_{coh}$ | $E_{form}$ | $\rho$ | $E_x$ | $E_g$ | Gap nature |
|---|---|---|---|---|---|---|---|---|---|---|---|
| TH-CN$_2$ | 4.18 | 5.78 | 1.48 | 1.47 | 1.46 | -5.26 | 0.87 | 6.30 | 340 | 4.57 | indirect |
| Tetrahex-C | 4.53 | 6.10 | 1.16 | 1.53 | 1.34 | -7.12 | 0.87 | | 288 | 2.64 | direct |
| Penta-CN$_2$ | 3.31 | 3.31 | 1.52 | 1.47 | 1.47 | -5.51 | 0.61 | 4.41 | 315 | 6.53 | indirect |
| Penta-graphene | 3.64 | 3.64 | 1.20 | 1.55 | 1.34 | -7.09 | 0.91 | | 264 | 3.25 | indirect |
| Graphene | 2.46 | 2.46 | 0.00 | 1.42 | 1.42 | -7.99 | 0* | | 348 | 0.00 | |

Covalent bonded C-N and N-N bonds are found in this 2D TH-CN$_2$ structure, which are supported by the atomic charge analysis using the Bader scheme [45]. The Bader charges on the C and N atoms are 2.8$e$ and 5.6$e$, respectively, implying charge transfer from C toward N atoms. In addition, N-N single bonds (i.e. $r_2$) are observed in the structure. The N-N single and N≡N triple bonds have bond energies of 160 kJ/mol and 954 kJ/mol [46], respectively This dramatic energy difference enables enormous energy stored in materials containing N-N single bonds. Therefore, such structure has great potential to be used as HEDM because the decomposition of the N-N single bonds into nitrogen molecules leads to a release of a large amount of energy. To estimate the amount of energy released from the decomposition of the compound, we further calculate the cohesive and formation energies of the material.

Cohesive energy of a solid is defined as the energy required to break the atoms of the solid into isolated atomic species. Therefore, the cohesive energy $E_{coh}$ per atom for a general carbon nitride compound $C_xN_y$ is computed using the following equation,

$$E_{coh} = \frac{E_{tot}(C_xN_y) - x\,E_C - y\,E_N}{x+y}, \tag{1}$$

where $E_{tot}$, $E_C$, $E_N$ are the total energy of the material, and the energies of an isolated carbon and nitrogen atoms, respectively. The formation energy $E_{form}$ per atom can then be obtained as follows [47,48],

$$E_{form} = E_{coh}(C_xN_y) - \frac{x}{x+y}\mu_C - \frac{y}{x+y}\mu_N, \text{ or } E_{form} = \frac{E_{tot}(C_xN_y) - x\,E(C) - y\,E(N)}{x+y}, \tag{2}$$

where $\mu_C$, $\mu_N$ are cohesive energy per atom of graphene and molecular N$_2$, respectively, $E(C)$, $E(N)$ are the total energies per atom for graphene and molecular N$_2$, respectively. Note that $\mu_C = E(C) - E_C$ and $\mu_N = E(N) - E_N$.

The calculated cohesive and formation energies are provided in Table 1. It is found that TH-CN$_2$ has slightly higher cohesive energy by 0.25 eV/atom, compared to penta-CN$_2$. It is opposite to their carbon counterparts since tetrahex-C is slightly energetically more favorable than penta-graphene (i.e. -7.12 versus -7.09 eV/atom). The formation energy of TH-CN$_2$ is found to be 0.87 eV/atom, corresponding to an energy density of 6.30 kJ/g. The energy density is calculated from the formation energy by converting energy unit of eV to kJ, combined with 1/3 and 2/3 of atomic mass values of C and N atoms, respectively. This energy density is 43% higher than that in penta-CN$_2$ [26], 83% higher than that 3.44 kJ/g in a recently predicted B$_3$N$_5$ compound [27].

The positive formation energy indicates that the structure of TH-CN$_2$ is metastable, similar to many other carbon nitrides [18,21,47,49–51], including $\beta$-C$_3$N$_4$ [52,53] and graphitic-C$_3$N$_4$ [54] which were already synthesized in lab. The new TH-CN$_2$ structure is proven to be thermodynamically stable according to the phonon spectrum calculations and AIMD. Its phonon spectrum and potential energy fluctuation with time are presented in Fig. 2. No imaginary frequency is observed in the phonon spectrum for the structure, indicating it is stable at low temperature. To further check the thermal stability of the structure at finite temperature, AIMD simulation is performed using a 4 × 4 supercell. There is no significant distortion of the structure after heating at 300 K for 6 picosecond with a time step of one femtosecond. The total potential energy fluctuation with time shows a stabilized magnitude in Fig. 2(b).

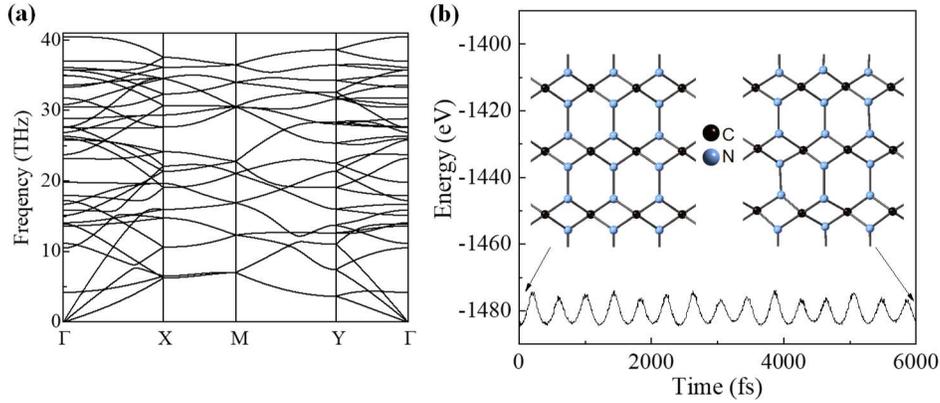

*Figure 2. Stability study of the 2D TH-CN₂ structure. (a) Phonon spectrum, (b) total potential energy (4 × 4 supercell) fluctuation with time during an ab initio molecular dynamics simulation at 300 K.*

We also explored the stability of the structure under mechanical strain. Starting with the fully relaxed 2D crystal structure of TH-CN$_2$, biaxial and uniaxial tensile strain up to 40% at an increment of 1% is applied in either the $x$ (**a** or zigzag) or $y$ (**b** or armchair) direction. The tensile strain is defined as,

$$\varepsilon = \frac{a - a_0}{a_0} \tag{3}$$

where $a$ and $a_0$ are the lattice constants of the strained and optimized structures, respectively. In the case of uniaxial strain applied in one direction, the lattice constant in the transverse direction is fully relaxed to ensure minimal stress in the transverse direction.

The phonon spectra of the material under various strain are shown in Fig. 3. No negative frequency is observed in the top row of the spectra. However, negative frequencies appear in the bottom row, indicating instability. The TH-CN$_2$ structure remains stable up to 10%, 13% and 5% for the uniaxial strain in the zigzag, armchair directions, and biaxial strain, respectively. However, the structure becomes unstable with higher strains, for example, 11%, 14% and 6% of strain in the zigzag, armchair and biaxial directions, respectively, since negative frequencies are introduced in their phonon spectrum as shown in Fig. 3 (d)-(f).

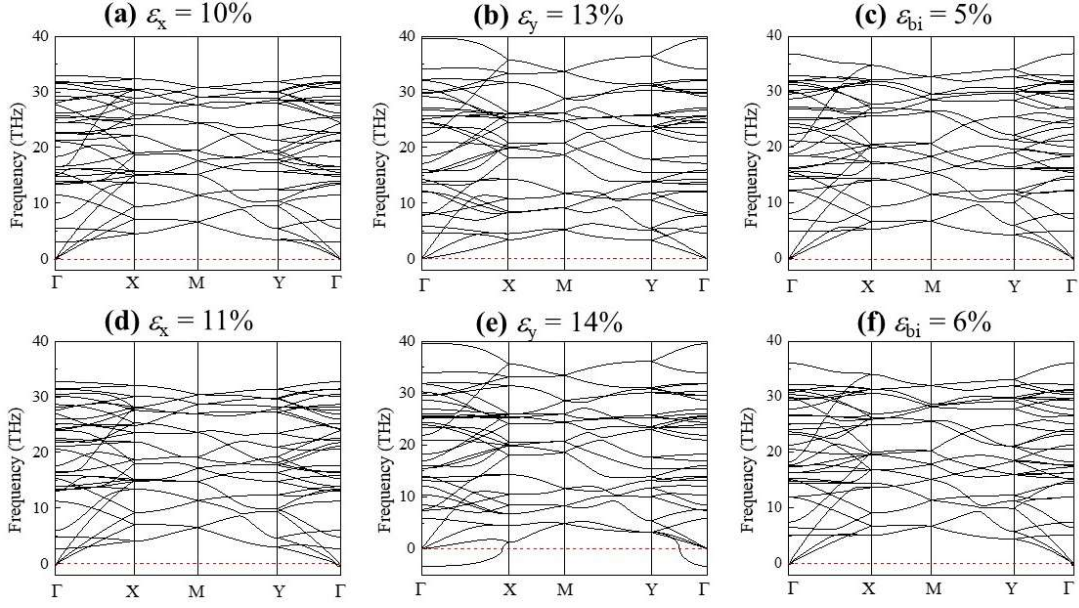

*Figure 3. Phonon spectra in the strained 2D TH-CN$_2$. (a) 10% uniaxial strain in the zigzag direction, (b) 13% uniaxial strain in the armchair direction, (c) 5% biaxial strain, (d) 11% uniaxial strain in the zigzag direction, (e) 14% uniaxial strain in the armchair direction, (f) 6% biaxial strain. The appearance of negative frequencies near Γ in the bottom rows indicates structural instability. The notations $\varepsilon_x/\varepsilon_y$, and $\varepsilon_{bi}$ on the top of figures represent uniaxial strain in the zigzag/armchair direction and biaxial strain, respectively.*

### 3.2. Mechanical properties

Exploration of strain-stress relation in a material can determine its ideal strength (the highest strength of a crystal at 0 K) [55,56] and critical strain (at which ideal strength reaches)[57]. To obtain the strain-stress relation in the 2D TH-CN$_2$ structure, uniaxial tensile strain along both $x$ (zigzag) and $y$ (armchair) directions and biaxial strain are applied to the material with an increment of 1% up to 40%. The result is given in Fig. 4 (a). The strain-stress relation is calculated using the method described in the references [58,59], which was designed for three dimensional (3D) material. For a 2D system, the stress calculated from the DFT has to be adjusted since the DFT reported stress is largely underestimated due to averaging force over vacuum space. To avoid this, the stress in this work adopts the force per unit length in the unit of N/m. The structure is found to be teared apart with 28% uniaxial strain applied in the $x$ axis or 35% biaxial strain. It is also found that the material is more ductile in the $y$ (armchair) direction, which is opposite to the case of tetrahex-C [43].

Since phonon instability occurs when strain is beyond 10%, 13%, and 5% in the $x$, $y$, and biaxial directions, respectively, solid symbols in Fig. 4(a) represent stable structures and hollow symbols for instable structure. Therefore, the ideal strength of TH-CN$_2$ is found to be 28.8 N/m and 19.6 N/m in the *x* and *y* direction, respectively. The strength is predicted to be 14.2 N/m for the biaxial strain. This ultrahigh strength of 28.8 N/m outperforms penta-graphene which demonstrates 23.5 N/m strength with 18% uniaxial strain in both zigzag and armchair directions [60]. However, it is lower than that of graphene and tetrahex-C [43]. For graphene, ideal strength is reported to be 36.7 N/m (40.4 N/m) in the zigzag (armchair) direction [56]. And for the tetrahex-C, the ideal strength is 38.3 N/m and 37.8 N/m in the zigzag and armchair directions, respectively [43].

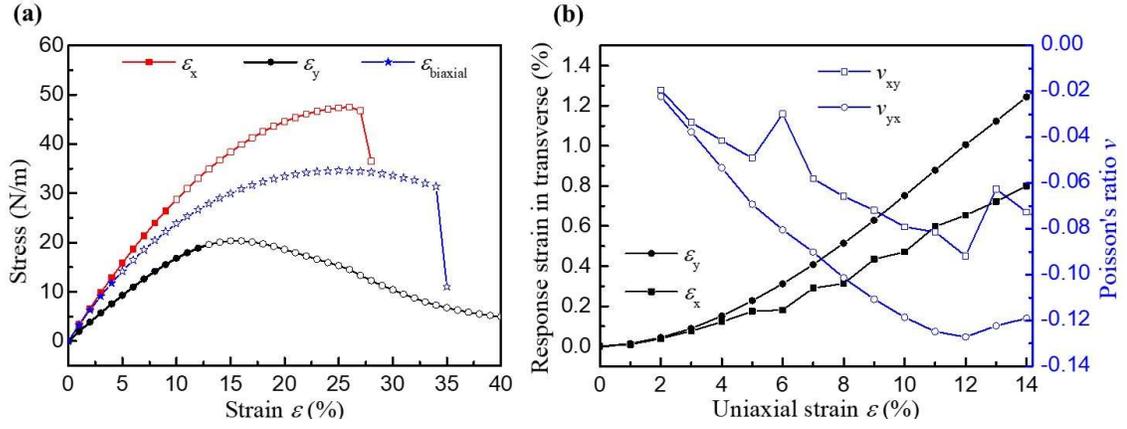

*Figure 4. (a) The strain-stress relation in the 2D TH-CN$_2$ for uniaxial strain applied in the zigzag, armchair, and biaxial directions, respectively. Phonon instability occurs when strain is beyond 10%, 13%, and 5% in the x, y, and biaxial directions with the corresponding strength 28.8, 19.6, and 14.2 N/m, respectively. The solid symbols and lines represent the stable structure and the hollow symbols indicating instable structures. (b) The response strain in the transverse direction and Poisson's ratio.*

As we mentioned previously, with uniaxial strain applied in one direction, the lattice constant in the transverse direction is fully relaxed to ensure minimal stress in the transverse direction. And the response strain in the transverse direction can be also calculated according to Eq. (3). Poisson's ratio can then be readily calculated according to its definition,

$$v = -\frac{\varepsilon_{transverse}}{\varepsilon_{axial}}, v_{xy} = -\frac{\varepsilon_y}{\varepsilon_x}, v_{yx} = -\frac{\varepsilon_x}{\varepsilon_y} \quad (4)$$

where $\varepsilon_{axial}$ and $\varepsilon_{transverse}$ are the applied axial strain and its response strain in the transverse direction, respectively. In order to depict the nonlinear lattice response for finite strain, Poisson's ratio is usually calculated using finite difference method as [61–63],

$$v = -\frac{d\,\varepsilon_{transverse}}{d\,\varepsilon_{axial}} \quad (5)$$

In our numerical calculations, Poisson's ratio is computed using the central finite difference method as [62],

$$v_{xy} = -\frac{\varepsilon_y^{j+1}-\varepsilon_y^{j-1}}{\varepsilon_x^{j+1}-\varepsilon_x^{j-1}}, v_{yx} = -\frac{\varepsilon_x^{j+1}-\varepsilon_x^{j-1}}{\varepsilon_y^{j+1}-\varepsilon_y^{j-1}}, \tag{6}$$

where the integer $j$ represents the strain increment number.

Fig. 4(b) presents the calculated response strain in the transverse direction and its corresponding Poisson's ratio in the TH-CN$_2$ structure. This structure demonstrates intrinsic in-plane negative Poisson's ratio with uniaxial strain holding up to the critical strain of 10% (13%) in the $x$ ($y$) direction. The Poisson's ratio is in the range of -0.02 ~ -0.12 for the uniaxial strain up to 13% in the armchair ($y$) direction. Similarly, the value is between -0.02 ~ -0.08 with strain up to 10% applied in the zigzag ($x$) direction. This demonstrates that the 2D TH-CN$_2$ structure possesses anisotropic feature.

To understand the intrinsic negative Poisson's ratio, we compared and analyzed the relaxed and strained TH-CN$_2$ structures and the results are shown in Fig. 5. The bond lengths $r_1$, $r_2$, buckling thickness $d$, bond angles $\alpha$, $\beta$ and $\gamma$ are denoted in Fig. 1(a). The dihedral angle $\phi_{1234}$ describes the angle between two intersecting planes where two neighboring hexagonal rings staying and $\phi_{2345}$ is the dihedral angle of neighboring hexagonal and tetragonal rings. Fig. 5 presents the change of each quantity relative to its original value in the relaxed structure as a function of uniaxial strain in the $x$ and $y$ directions, respectively. From Fig. 5(a)(c)(e), it is found that when the structure is under uniaxial strain in the zigzag direction, the bond length $r_1$ and bond angle $\gamma$ experience dominant changes, having 4.6% and 6.8% increase (for the case of $\varepsilon_x = 10\%$), respectively, compared to the relaxed structure. The bond length $r_1$ describes the C-N bonds (see Fig. 1(a)) which are tilted from the $x$-axis with an apparent $y$-projection. With the uniaxial strain in the $x$-direction, the elongation in $r_1$ inevitably leads to the extension of the crystal lattice the $y$-axis, resulting in the negative Poisson's ratio.

On the other hand, for the case of strain in the armchair direction in Fig. 5(b) (d) (f), the major changes are from the squeeze of the buckling thickness $d$ (11.9% reduction for the case of $\varepsilon_y = 12\%$), the elongation of bond length $r_2$ (10.4% up), and the decrease of bond angle $\alpha$ (6.7% down). The combined effect of the squeezing of $d$ and lengthening of $r_1$ serve as the primary factors for the negative Poisson's ratio.

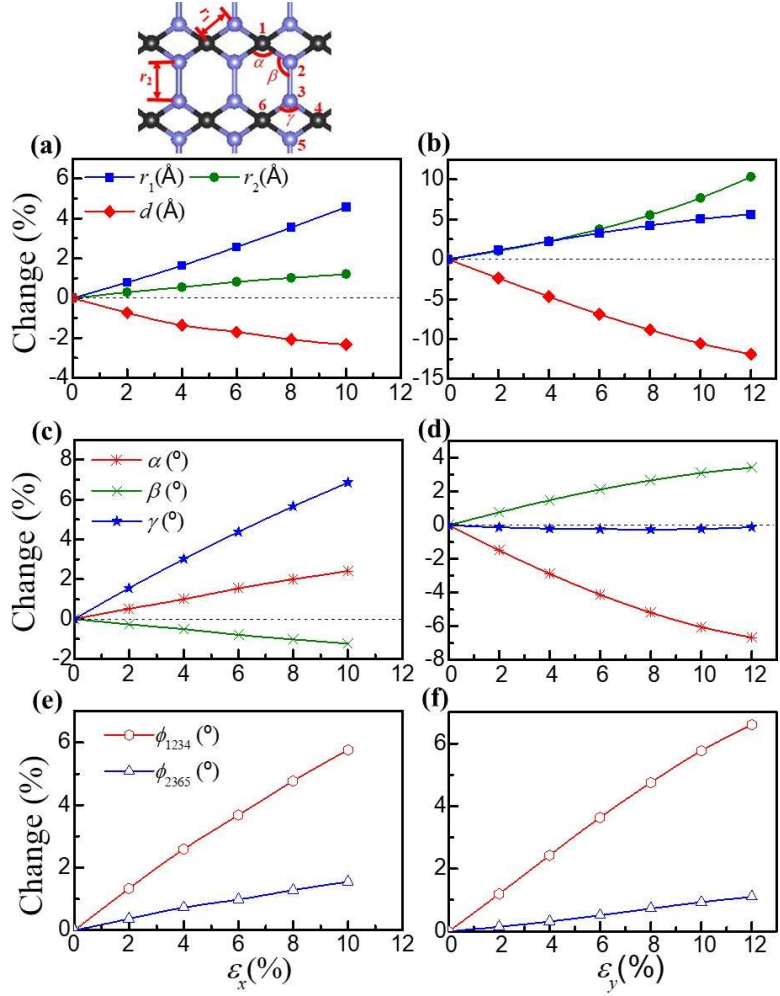

*Figure 5. Structural change of TH-CN$_2$ under uniaxial strain applied in the (a) (c) (e) zigzag and (b) (d) (f) armchair direction. (a) (b) The bond lengths, buckling thickness d, (c) (d) bond angles, and (e) (f) dihedral angles. The structural parameters are denoted in Figure 1(a) and displayed on the top of the figure. Vertical axis represents the change relative to their original values in the relaxed structure.*

To calculate elastic stiffness constants and various moduli, the energy surface of the 2D TH-CN$_2$ structure is scanned in the small strain range $-0.6\% < \varepsilon_{xx} < +0.6\%$, $-0.6\% < \varepsilon_{yy} < +0.6\%$, and $-0.6\% < \varepsilon_{xy} < +0.6\%$. The strain energy is defined as the energy difference between the strained and relaxed structures,

$$E_s = E(\varepsilon) - E_0 \tag{7}$$

where $E(\varepsilon)$ and $E_0$ are the total energy of strained and relaxed structures, respectively. The calculated strain energy is then fitted parabolically using the following equation,

$$E_s = a_1 \varepsilon_{xx}^2 + a_2 \varepsilon_{yy}^2 + a_3 \varepsilon_{xx} \varepsilon_{yy} + a_4 \varepsilon_{xy}^2 \tag{8}$$

to determine the coefficients $a_i$, and the elastic stiffness constants are readily calculated as,

$$C_{ij} = \frac{1}{A_0}\left(\frac{\partial E_s^2}{\partial \varepsilon_i \varepsilon_j}\right), \tag{9}$$

where $i, j = xx, yy$, or $xy$, $A_0$ is the area of the simulation cell in the $xy$ plane. The Young's and shear moduli for a 2D system can be derived as a function of $a_i$ [43,57],

$$E_x = \frac{4a_1a_2 - a_3^2}{2a_2 A_0}, E_y = \frac{4a_1a_2 - a_3^2}{2a_1 A_0}, G_{xy} = \frac{2a_4}{A_0}. \tag{10}$$

The elastic stiffness constants in TH-CN$_2$ are predicted to be $C_{11}$ = 340 N/m, $C_{12}$ = -1.5 N/m, $C_{22}$ = 196 N/m, $C_{33}$ = 87 N/m. The Young's moduli are $E_x$ = 340 N/m and $E_y$ = 196 N/m in the zigzag and armchair directions, respectively. This tremendously large modulus value in the zigzag direction is comparable to that in graphene and larger than that in penta-CN$_2$, penta-graphene and tetrahex-C, as shown in Table 1. The shear modulus is predicted to be $G_{xy}$ = 87 N/m.

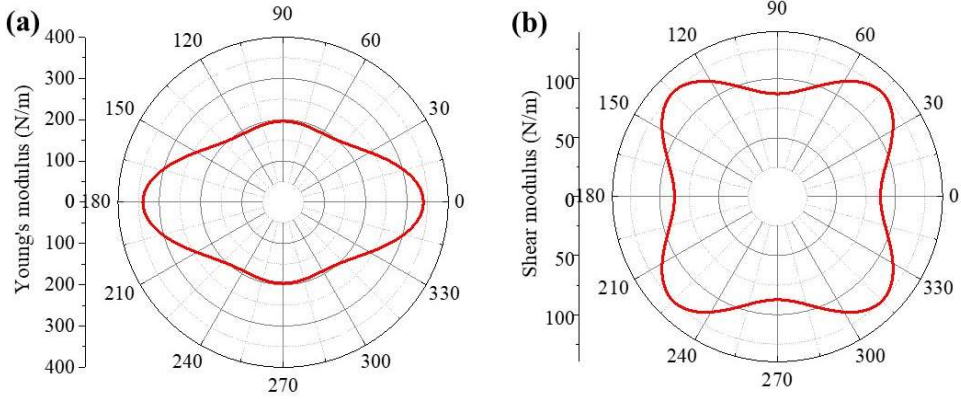

Figure 6. The directional dependence of (a) Young's modulus and (b) shear modulus in TH-CN$_2$.

In addition, the Young's and shear moduli along an arbitrary direction can be calculated using the following equations [57],

$$\frac{1}{E_\varphi} = S_{11}\cos^4\varphi + (2S_{12} + S_{66})\cos^2\varphi\sin^2\varphi + S_{22}\sin^4\varphi \tag{11}$$

$$\frac{1}{G_\varphi} = S_{33}(\sin^4\varphi + \cos^4\varphi) + 4\left(S_{11} - 2S_{12} + S_{22} - \frac{1}{2}S_{33}\right)\cos^2\varphi\sin^2\varphi \tag{12}$$

where $\varphi \in [0, 2\pi]$ is the angle of an arbitrary direction from the $+x$ axis, $E_\varphi$ and $G_\varphi$ are the Young's and shear moduli, respectively, along that particular direction, $S_{ij}$ are elastic compliance constants, which are correlated to the elastic stiffness constants as following,

$$S_{11} = \frac{C_{22}}{C_{11}C_{22} - C_{12}^2}, S_{22} = \frac{C_{11}}{C_{11}C_{22} - C_{12}^2}, S_{12} = -\frac{C_{12}}{C_{11}C_{22} - C_{12}^2}, S_{33} = \frac{1}{C_{33}} \tag{13}$$

The direction dependence of the Young's and shear moduli are presented in Fig. 6. The maximal Young's modulus is along the $x$ (zigzag) direction with a value of 340 N/m, while a minimum of 196 N/m is along the $y$ (armchair) direction. However, it is a different story for the shear modulus. The maximal shear modulus is along the [11] direction with a value of 125 N/m and the minimum occurs in the and $x(y)$-directions (87 N/m). This again demonstrates a strong anisotropicity in the 2D TH-$CN_2$ structure.

### 3.3. Electronic properties

The electronic band structure of TH-$CN_2$ is presented in Fig. 7 (a) using the hybrid HSE functional. It shows that TH-$CN_2$ possesses an indirect band gap with a value of 4.57 eV (the PBE functional predicted band gap is 3.04 eV). This gap is smaller than that 6.53 eV in penta-$CN_2$, but larger than the gaps in tetrahex-C and penta-graphene, as shown in Table 1.

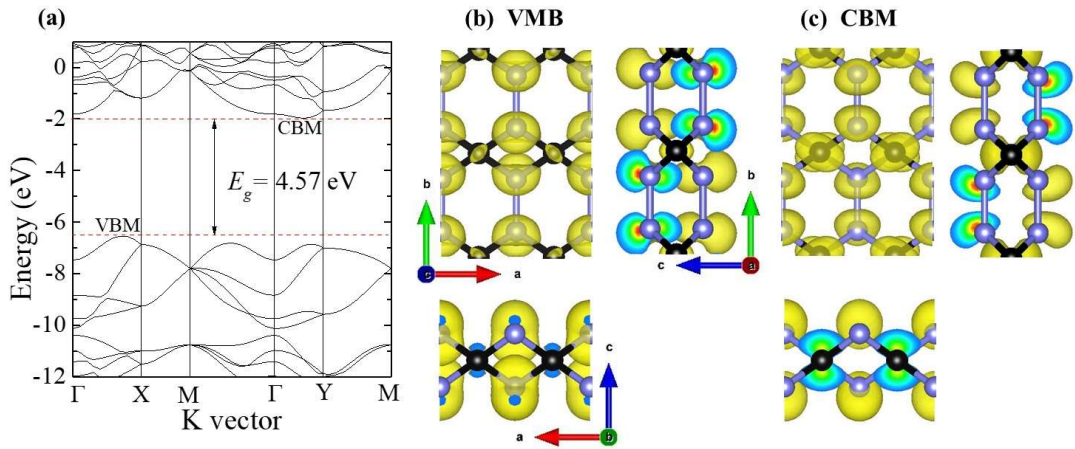

Figure 7. (a) The electronic band structure of TH-$CN_2$ predicted by the hybrid HSE methods. The indirect band gap is 4.57 eV. Energy is referenced to vacuum. (b) The electron orbital contour plots of the (b) VBM and (c) CBM. VBM is dominated by $p_z$ orbital on N atoms. CBM is mainly contributed by $p_z$ orbital on C and $s$-$p_y$ hybridization on N atoms. The isosurface value for the charge density is set to be 0.06 $e/Bohr^3$.

The valence band maximum (VBM) is found to be along the wave vector direction Γ-X, while the conduction band minimum (CBM) locates along the high symmetry line of Γ-Y. The electron orbital contour plots of the VBM and CBM are shown in Fig. 7(b) (c). Through an analysis of the *spd*-orbital site projection of the VBM and CBM, it is found that the VBM is dominantly contributed by the $p_z$ orbital on the N atoms, while the CBM is dominated by the $p_z$ orbital on the C atoms and the $s$-$p_y$ hybridization on the N atoms.

Strain engineering of material properties is a commonly used technique in science and technology [43,57,64–73]. We find that the electronic band structure of TH-$CN_2$ can be effectively modified by uniaxial and biaxial strains. Fig. 8 shows the variation of the band structure in TH-$CN_2$ with strain. Black dashed lines are for the relaxed structure, while the red solid lines are for the strained system. It is clear that all uniaxial and biaxial strains can tune the band structure and reduce its band gap. The band gap variation with strain is given in Fig. 8(d), which presents the gaps predicted from both HSE and PBE functionals. As mentioned before, since phonon spectrum calculation suggests that the TH-CN$_2$ structure maintain stable in the range up to 10% (13%) and 5% for the uniaxial strain in the $x(y)$ direction and biaxial strain, Fig. 8(d) only plots the band gap variations within that strain range with phonon stability. All of the predicted band gaps are indirect and there is no indirect-direct band gap transition observed. Fig. 8(d) also suggests that the uniaxial strain in the $y$ direction is the most effective to tune the band gap and the axial strain in the $x$ direction is the least. This may be resulted from the fact that $\varepsilon_y$ strain is more effective to distort the 2D structure, compared to $\varepsilon_x$ strain as demonstrated in Fig. 5.

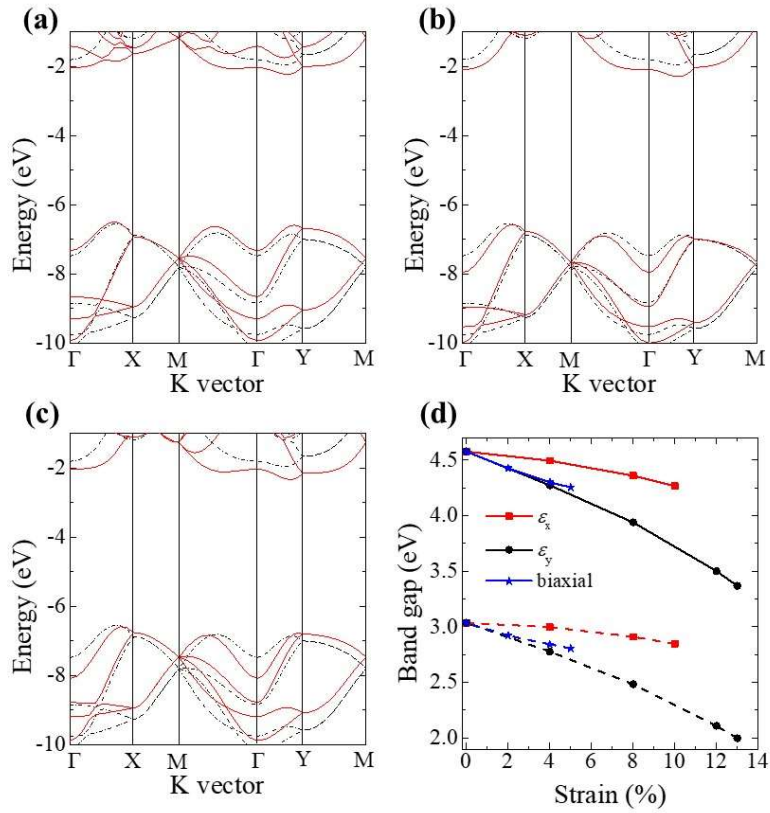

*Figure 8. Variation of electronic band structure with strain in TH-CN$_2$. The HSE predicted band structure comparison between the relaxed TH-CN$_2$ and the one with strain (a) $\varepsilon_x = 10\%$, (b) $\varepsilon_y = 4\%$, (c) $\varepsilon_{bi} = 5\%$. Black dashed and red solid lines are for the relaxed and strained structures, respectively. Energy is referenced to vacuum. (d) The HSE (solid lines) and PBE (dashed lines) predicted band gap as a function of strain. All band gaps are indirect in the strain range with phonon stability.*

### 3.4. Double layers

In addition to a single layer of TH-CN$_2$, we also explored the structures of double layers. There are several scenarios to generate double layers. One is just topping a layer on the other, called AA-stacking as shown in Fig. 9 (a)-(c). The other is similar but flips over the top layer and named as AB-stacking, shown in Fig. 9(a) (b) (d). Our calculations verify that the AA stacking has slightly lower energy that that of AB stacking. The third scenario to create double layers is follows, starting with the AA stacking, shift the top layer along the *y* (armchair) direction by a quarter of its lattice constant. The resulting structure named as AA-shifted is shown in Fig. 9(e) (f). If the top layer shifted along the *y* (*x*) direction by a half of its lattice constants, it will give the AB-stacking.

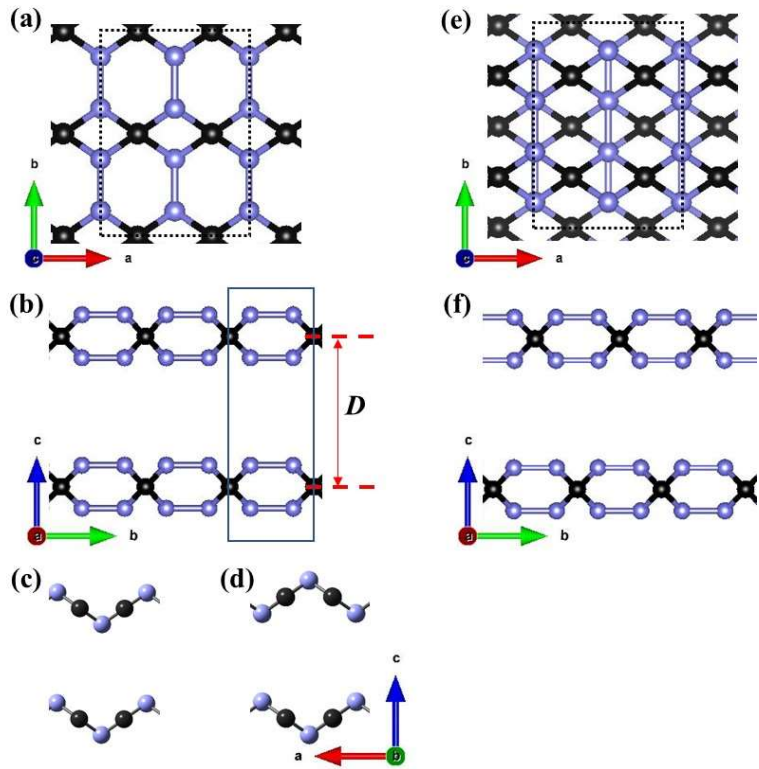

*Figure 9. Double-layer TH-CN$_2$ structures, (a)-(d) AA and AB stacking, (c) AA stacking, (d) AB stacking, (e)-(f) AA-shifted.*

To improve the description of Van der Waals interaction between layers, two additional functionals PBEsol [38] and DFT-D3 [39] are implemented in addition to the PBE functional. All three functionals are used to calculate the structure and electronic properties of the double layers. The calculated lattice constants *a* and *b*, interlayer distance *D*, total energies, and band gaps of the three scenarios of double layers are listed in Table 2. It is found that the results obtained from the PBE functional are very close to that from DFT-D3. Therefore, we only list the structural results from PBE and PBEsol in the table.

*Table 2. Calculated lattice constants a, b and interlayer distance D in unit of Å, total energies in meV relative to the AA-shifted case, predicted band gap in eV for the double-layer TH-CN$_2$ structures.*

| Double layers | PBE | | | PBEsol | | | Energy | | Gap | | |
|---|---|---|---|---|---|---|---|---|---|---|---|
| | *a* | *b* | *D* | *a* | *b* | *D* | PBE | PBEsol | PBE | PBEsol | HSE06 |
| AA | 4.18 | 5.78 | 5.18 | 4.17 | 5.76 | 4.93 | 5 | 2 | 2.95 | 2.94 | 4.48 |
| AB | 4.18 | 5.78 | 5.44 | 4.17 | 5.76 | 5.44 | 15 | 18 | 2.95 | 2.97 | 4.48 |
| AA-shifted | 4.18 | 5.78 | 5.18 | 4.17 | 5.76 | 5.18 | 0 | 0 | 2.99 | 3.02 | 4.53 |

Among three different scenarios of generating double layers, it is found that AA-shifted case is the most energetically favorable with a total energy of -186.168 eV predicted by the PBE functional. And the structures with the AA and AB stacking have total energies of -186.163 and -186.153 eV, respectively, which are 5 and 15 meV higher than that of AA-shifted case. This energy sequence in which AA-shifted structure has the lowest energy, followed by AA and AB stacking, is further validated by the PBEsol method as shown by the data in Table 2. Regarding to the interlayer distance *D*, the AB stacking possesses the largest distance due to its special arrangement of layers with the N-N bonds aligned facing each other. The PBE functional predicts the same interlayer distance for the AA stacking and shifted cases. However, PBEsol differentiates that by its improved description of interlayer interaction. It predicts a greater interlayer distance for the AA-shifted case than that in AA-stacking. The large values of the interlayer distance suggests that no chemical bonds form between layers. We even used small interlayer distance in our initial structural setting, but the relaxation from the DFT calculation pushes two layers apart to reach the interlayer distance reported in Table 2. Negligible differences in the lattice constants *a* and *b* are observed for the three types of double layers, regardless of the functionals being used.

The AA-shifted case has the largest band gap 4.53 eV among the three scenarios, which is slightly lower than the value of single layer (4.57 eV). The HSE calculated band structure for the AA-shifted case is presented in Fig. 10. It possesses an indirect band gap with the VBM along Γ-X and the CBM along Γ-Y, similar to the case of single layer. It can be found in Fig. 10 that some energy bands are degenerate (e.g. VBM) while others show release of degeneracy but with very little splitting of energy (e.g. CBM). This is no surprise due to its double-layer structure with weak interlayer interaction.

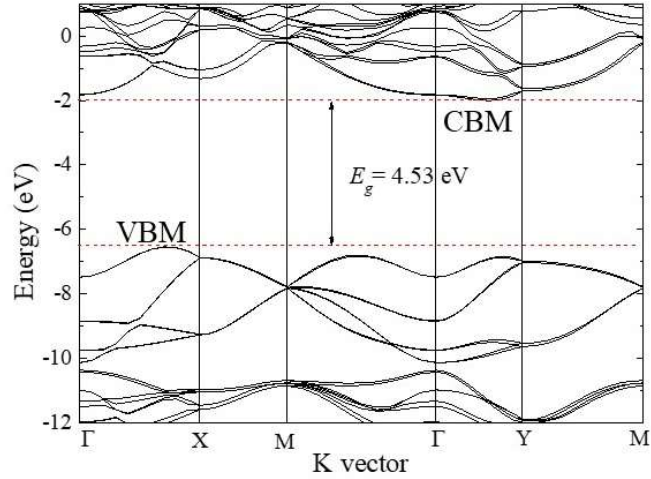

*Figure 10. The HSE predicted electronic band structure of the double-layer TH-CN$_2$, the AA-shifted case.*

## 4. Summary

Using first-principles DFT calculations and AIMD, we propose a new 2D CN$_2$ structure, named TH-CN$_2$. We find this structure is thermodynamically stable. It contains single N-N bonds with 67% nitrogen concentration and high energy density for potential applications in HEDM. It possesses exotic mechanical properties including ultrahigh strength and negative Poisson's ratio. It exhibits strong anisotropicity and its Young's modulus along the zigzag direction reaches 340 N/m, close to that in graphene (348 N/m). The material has a wide indirect band gap of 4.57 eV and the band gap can be effectively tuned in the range of 3.37 eV – 4.57 eV through the application of mechanical strain. Double-layer structure is also explored and the energetically most favorable scenario is proposed. The high energy density and strength, strong moduli, negative Poisson's ratio, and wide band gap in this 2D TH-CN$_2$ structure may have potential applications in energy storage, nanomechanics and nanoelectronics.

**Acknowledgement**

This work is financially supported by the Natural Science Foundation of China (Grant No.: 11965005), Natural Science Basic Research plan in Shaanxi Province of China (Grant Nos.: 2020JM-186), the 111 Project (B17035), and the Fundamental Research Funds for the Central Universities. The authors thank Arizona State University Advanced Computing Center for